\documentclass{sig-alternate-2013}

\usepackage{booktabs} 

\usepackage{colortbl}

\usepackage{amsmath}
\usepackage{amsfonts}
\usepackage{graphicx}
\usepackage{subfigure}
\usepackage{url}
\usepackage{paralist} 
\usepackage{listings} 
\usepackage{lipsum}
\usepackage{cuted}
\usepackage{algorithmic}
\usepackage{algorithm}
\usepackage{rotating}
\usepackage{pifont}
\usepackage{xspace}
\newcommand{\cmark}{\ding{51}}%

\usepackage{xcolor}

\usepackage{empheq}

\newcommand{\nop}[1]{{}}

\newcommand{\xhdr}[1]{\vspace{1.7mm}\noindent{{\bf #1.}}}

\definecolor{speagle}{RGB}{154,32,0}
\definecolor{bad}{RGB}{204,121,167}
\definecolor{birdnest}{RGB}{0,230,163}
\definecolor{trustiness}{RGB}{230,106,0}
\definecolor{fairjudge}{RGB}{0,115,179}
\definecolor{fairjudgep}{RGB}{230,163,0}
\definecolor{maroon}{HTML}{800000}
\definecolor{sb}{HTML}{DDA0DD}

\newcommand{\highlight}[1]{\colorbox{gray!25}{$\displaystyle#1$}}
\newcommand{\hl}[1]{\colorbox{yellow!50}{$\displaystyle#1$}}

\usepackage{tikz}
\newtheorem{example}{Example}
\newtheorem{lemma}{Lemma}
\newtheorem{theorem}{Theorem}
\newtheorem{corollary}{Corollary}

\newcommand{\hide}[1]{}

\newcommand{\bit}{\begin{compactitem}}
\newcommand{\eit}{\end{compactitem}}
\newcommand{\ben}{\begin{compactenum}}
\newcommand{\een}{\end{compactenum}}

\newcommand{\methodall}{\textsc{FairJudge}}

\begin{document}

\title{FairJudge: Trustworthy User Prediction in Rating Platforms}

\numberofauthors{6}
\author{
\alignauthor
\hspace{3mm}
Srijan Kumar\\
       \affaddr{University of Maryland}\\
       \email{srijan@cs.umd.edu}
\alignauthor
Bryan Hooi \\
       \affaddr{Carnegie Mellon University}\\
       \email{bhooi@cs.cmu.edu}
\alignauthor
Disha Makhija \\
       \affaddr{Flipkart, India}\\
       \email{disha.makhija@flipkart.com}
\and
\alignauthor
Mohit Kumar \\
       \affaddr{Flipkart, India}\\
       \email{k.mohit@flipkart.com}
\alignauthor
Christos Faloutsos \\
       \affaddr{Carnegie Mellon University}\\
       \email{christos@cs.cmu.edu}
\alignauthor
V.S. Subrahmanian \\
       \affaddr{University of Maryland}\\
       \email{vs@cs.umd.edu}
}

\maketitle

\begin{abstract}
Rating platforms enable large-scale collection of user opinion about items (products, other users, etc.).
However, many untrustworthy users 
give fraudulent ratings for excessive monetary gains.
In the paper, we present \methodall, a system to identify such fraudulent users.
We propose three metrics:
(i) the \emph{fairness of a user} that quantifies how trustworthy the user is in rating the products,
(ii) the \emph{reliability of a rating} that measures how reliable the rating is, and
(iii) the \emph{goodness of a product} that measures the quality of the product. 
Intuitively, a user is fair if it provides reliable ratings that are close to the goodness of the product.
We formulate a mutually recursive definition of these metrics, and further address \textit{cold start problems} and incorporate \textit{behavioral properties} of users and products in the formulation.
We propose an iterative algorithm, \methodall, to predict the values of the three metrics.
We prove that \methodall\ is guaranteed to converge in a bounded number of iterations, with linear time complexity. 
By conducting five different experiments on five rating platforms, we show that \methodall\ significantly outperforms nine existing algorithms in predicting fair and unfair users.
We reported the 100 most unfair users 
in the Flipkart network to their review fraud investigators, and 80
users were correctly identified (80\% accuracy). 
The \methodall\ algorithm is already being deployed at Flipkart.
\end{abstract}

\section{Introduction}

\label{sec:intro}
Consumer generated ratings are now an essential part of several platforms. For instance, users on Yelp and TripAdvisor rate restaurants, hotels, attractions, and various types of service offerings. Every major online marketplace (eBay, Amazon, Flipkart) uses online ratings as a way of recognizing good products and rewarding honest/good behavior by vendors. Because buyers frequently look at reviews before buying a product or using a vendor, there is a huge incentive for unethical entities to give fraudulent ratings to 
compromise the integrity of these consumer generated rating systems \cite{lappas2016impact,jindal2008opinion,wang2012bonus,hooi2016fraudar}. The goal of this paper is to identify such unfair users.

In this paper, we present three novel metrics to quantify the trustworthiness of users and reviews, and the quality of products, building on our prior work~\cite{kumar2016wsn}.
We model user-to-item ratings with timestamps as a bipartite graph. For instance, on an online marketplace such as Amazon, a user $u$ rates a product $p$ with a rating $(u,p)$. Each user has an intrinsic level of fairness $F(u)$, each product $p$ has an intrinsic goodness $G(p)$ (measuring its quality), and each rating $(u,p)$ has an intrinsic reliability $R(u,p)$. 
Intuitively, a fair user should give ratings that are close to the goodness of the product, and good products should get highly positive reliable ratings. Clearly, these $F(u),G(p),R(u,p)$ metrics are all inter-related, and we propose three mutually-recursive equations to model them. 

However, it is not possible to predict the true trustworthiness of users that have only given a few ratings. For example, users with only a few ratings, all of which are highly accurate, can be a fraudulent shill account building initial reputation~\cite{hooi2016fraudar} or it can be a genuine user. Similarly, the true quality of products that have received only a few ratings is also uncertain~\cite{lim2010detecting,trueview}. 
We propose a Bayesian solution to to address these \emph{cold start problems} in our formulation by incorporating priors to users' fairness and products' goodness scores.

Additionally, the rating behavior is often very indicative of their nature. For instance, unusually rapid or regular behavior has been associated with fraudulent entities, such as fake accounts, sybils and bots~\cite{viswanath2014towards,li2017bimodal,birdnest,viswanath2015strength}. Similarly, unusually bursty ratings received by a product may be indicative of fake reviews~\cite{wu2010merging}. Therefore, we propose a Bayesian technique to incorporate users' and products' rating behavior in the formulation, by penalizing unusual behavior~\cite{birdnest}. 

Combining the network, cold start treatment and behavioral properties together, we present the \methodall\ formulation and an iterative algorithm to find the fairness, goodness and reliability scores of all entities together.
We prove that \methodall\ has linear time complexity and it is guaranteed to converge in a bounded number of iterations.

\begin{figure}[t]
\centering
		\subfigure[]{
        \includegraphics[width=0.20\textwidth, clip=true, trim= 2.3mm 2.2mm 2.6mm 2.2mm]{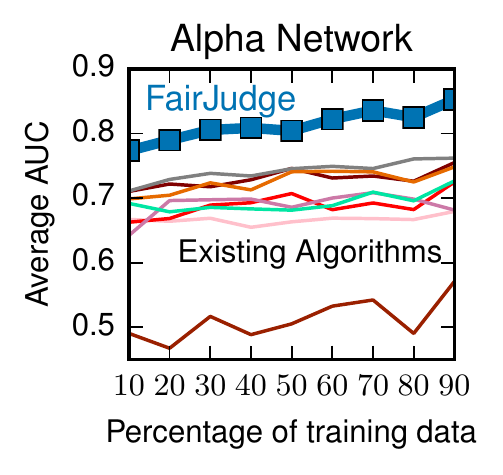}
        } 
        \subfigure[]{
        \includegraphics[width=0.22\textwidth, clip=true, trim= 1.6cm 0.90cm 1.6cm 1.9cm]{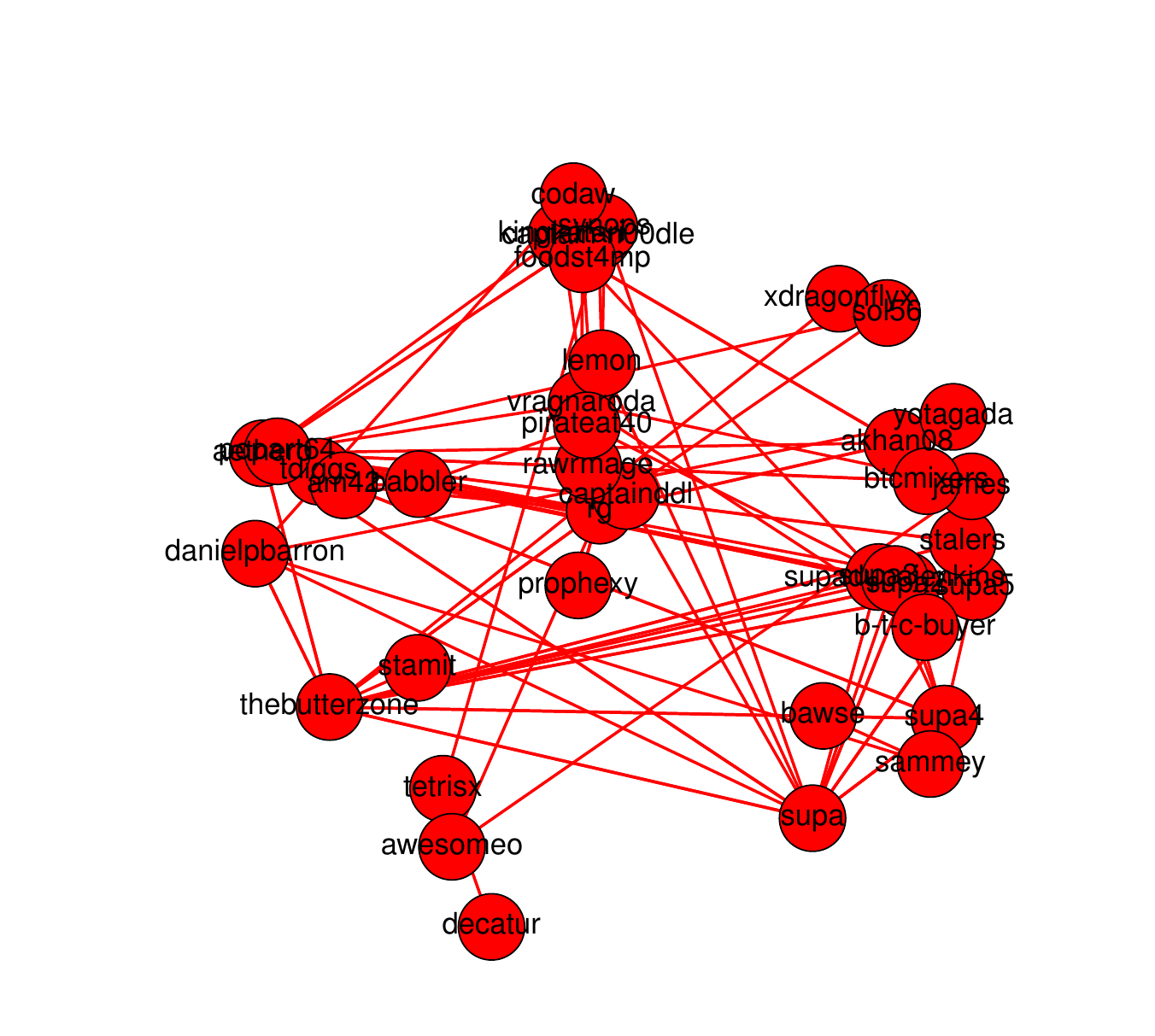}
        }  
		\vspace{-2mm}
    \caption{\small (a) The proposed algorithm, \methodall, consistently performs the best, by having the highest AUC, in predicting fair and unfair users with varying percentage of training labels. (b) \methodall\ discovered a bot-net of 40 confirmed shills of one user, rating each other positively.} 
    		\vspace{-6mm}
    \label{fig:crownjewel}
\end{figure}

How well does \methodall\ work? We conduct extensive experiments using 5 real-world data sets -- two Bitcoin user trust networks, Epinions, Amazon and Flipkart, India's biggest online marketplace.
With the help of \underline{five experiments}, we show that \methodall\ outperforms several existing methods~\cite{fraudeagle,birdnest,bad,speagle,icdm2011,mukherjee2013yelp,mukherjee2013spotting,lim2010detecting} in predicting fraudulent users.
Specifically, in an unsupervised setting, \methodall\ has the best or second best average precision in nine out of ten cases.
Across two supervised settings, \methodall\ has the highest AUC $\geq 0.85$ across all five datasets. It consistently performs the best as the percentage of training data is varied between 10 and 90, as shown for the Alpha network in Figure~\ref{fig:crownjewel}(a). 
Further, we experimentally show that both the cold start treatment and behavior properties improve the performance of \methodall\ algorithm, and incorporating both together performs even better. 

\methodall\ is practically very useful. We reported the 100 most unfair users as predicted by \methodall\ in the Flipkart online marketplace. Review fraud investigators at Flipkart studied our recommendations and confirmed 80 of them were unfair, 
presenting a validation of the utility of \methodall\ in identifying real-world review fraud.
In fact, \textit{\methodall\ is already being deployed at Flipkart.}
On the Bitcoin Alpha network, using \methodall, we discovered a botnet-like structure of 40 unfair users that rate each other positively (Figure~\ref{fig:crownjewel}(b)), which are confirmed shills of a single account.

Overall, the paper makes the following contributions:\\
$\bullet$ \textbf{Algorithm}: We propose three novel metrics called fairness, goodness and reliability to rank users, products and ratings, respectively. We propose Bayesian approaches to address cold start problems and incorporate behavioral properties. We propose the \methodall\ algorithm to iteratively compute these metrics.\\
$\bullet$ \textbf{Theoretical guarantees}: We prove that \methodall\ algorithm is guaranteed to converge in a bounded number of iterations, and it has linear time complexity. \\ 
$\bullet$ \textbf{Effectiveness}: We show that \methodall\ outperforms nine existing algorithms in identifying fair and unfair users, conducted via five experiments on five rating networks. 

\section{Related Work}
\label{sec:background}

Existing works in rating fraud detection can be categorized into network-based and behavior-based algorithms:\\
\textbf{Network-based fraud detection} algorithms are based on iterative learning, belief propagation,  and node ranking techniques.
Similar to our proposed \methodall\ algorithm, \cite{icdm2011,bad,li2012robust} develop iterative algorithms that jointly assign scores in the rating networks based on consensus of ratings - \cite{icdm2011} scores each user, review and product, and \cite{bad} scores each user and product.
FraudEagle~\cite{fraudeagle} is a belief propagation model to rank users, which assumes fraudsters rate good products poorly and bad products positively, and vice-versa for honest users. 
Random-walk based algorithms have been developed to detect trolls~\cite{wu2016troll} and link farming from collusion on Twitter~\cite{ghosh2012understanding}.
\cite{jiang2014catchsync, ye2015discovering, li2012robust} identify group of fraudsters based on local neighborhood of the users.
A \textit{survey} on network-based fraud detection can be found in \cite{DBLP:journals/datamine/AkogluTK15}. 

\textbf{Behavioral fraud detection} algorithms are often feature-based.
Consensus based features have been proposed in ~\cite{mukherjee2013yelp, lim2010detecting} -- our proposed goodness metric is also inspired by consensus or `wisdom of crowds'.
Commonly used features are derived from timestamps~\cite{xie2012review, wu2010merging, trueview} and review text~\cite{sun2013synthetic, fayazi2015uncovering, sandulescu2015detecting}.
SpEagle~\cite{speagle} extends FraudEagle~\cite{fraudeagle} to incorporate behavior features. 
BIRDNEST~\cite{birdnest} creates a Bayesian model to estimate the belief of each user's deviation in rating behavior from global expected behavior.
\cite{jiang2014catchsync, chen2013battling, viswanath2015strength} study coordinated spam behavior of multiple users.
A \textit{survey} on behavior based algorithms can be found in \cite{jiang2016suspicious}. 

Our proposed \methodall\ algorithm combines both network and behavior based algorithms, along with Bayesian solution to cold start problems. Since review text is not always available, for e.g. in a subset of our datasets, we only focus on ratings and timestamps.
\methodall\ provides theoretical guarantees and does not require any user inputs.
Table~\ref{tab:salesman} compares \methodall\ to the closest existing algorithms, which do not always satisfy all the desirable properties.

\begin{table}[t]
\small
\centering
\begin{tabular}{c||cccccc||c}
 & \begin{turn}{90}BIRDNEST\cite{birdnest}\end{turn} & \begin{turn}{90}Trustiness\cite{icdm2011}\end{turn} & \begin{turn}{90}BAD\cite{bad}\end{turn} & \begin{turn}{90}FraudEagle\cite{fraudeagle}\end{turn} & \begin{turn}{90}SpEagle\cite{speagle}\end{turn} & \begin{turn}{90}Others~\cite{mukherjee2013yelp, mukherjee2013spotting, lim2010detecting}\end{turn}   & \begin{turn}{90}\methodall\ \end{turn}  \\\hline
Uses network information & \cmark & \cmark & \cmark & \cmark & \cmark &  & \cmark \\
Uses behavior properties & \cmark & & & & \cmark & \cmark & \cmark\\
Prior free & & & \cmark &  &  & \cmark & \cmark \\
Theoretical Guarantees & & & \cmark & & & & \cmark \\\hline
\end{tabular}
\caption{Proposed algorithm \methodall\ satisfies all desirable properties.}
		\vspace{-4mm}
\label{tab:salesman}
\end{table}

\section{FairJudge Formulation}
\label{sec:formulation}
In this section, we present the \methodall\ algorithm that jointly models the rating network and behavioral properties.
We first present our three novel metrics --- Fairness, Reliability and
Goodness --- which measure intrinsic properties of users, ratings and products, respectively, building on our prior work~\cite{kumar2016wsn}. 
We show how to incorporate Bayesian priors to address user and product cold start problems, 
and how to incorporate behavioral features of the users and products. 
We then prove that our algorithm have several desirable theoretical guarantees.

\noindent\textbf{Prerequisites.}
We model the rating network as a bipartite network where user $u$ gives a rating $(u,p)$ to product $p$. Let the rating score be represented as \texttt{score(u,p)}.
Let $\mathcal{U}, \mathcal{R}$ and $\mathcal{P}$ represent the set of all users, ratings and products, respectively, in a given bipartite network. 
We assume that all rating scores are scaled to be between -1 and +1, i.e. $\texttt{score(u,p)} \in [-1, 1] \forall (u,p) \in \mathcal{R}$. 
Let, \texttt{Out(u)} be the set of ratings given by user $u$ and \texttt{In(p)} be the set of ratings received by product $p$. So, $|\texttt{Out(u)}|$ and $|\texttt{In(p)}|$ represents their respective counts.

\subsection{Fairness, Goodness and Reliability} \label{sec:assumptions}
Users, ratings and products have the following characteristics: 

$\bullet$ Users vary in {\it \textbf{fairness}}. 
Fair users rate products without bias, i.e. they give high scores to high quality products, and low scores to bad products.
On the other hand, users who frequently deviate from the above behavior are `unfair'. For example, fraudulent users often create multiple accounts to boost ratings of unpopular products and bad-mouth good products of their competitors~\cite{viswanath2015strength,hooi2016fraudar}.
Hence, these fraudsters should have low fairness scores.
The fairness $F(u)$ of a user $u$ lies in the $[0,1]$ interval $\forall u \in \mathcal{U}$. 0 denotes a 100\% untrustworthy user, while 1 denotes a 100\% trustworthy user.

$\bullet$ Products vary in terms of their quality, 
which we measure by a metric called {\it \textbf{goodness}}.
The quality of a product determines how it would be rated by a fair user. 
Intuitively, a good product would get several high positive ratings 
from fair users, and a bad product would receive high negative ratings from
fair users.
The goodness $G(p)$ of a product $p$ ranges from $-1$ (a very low quality product) to $+1$ (a very high quality product) $\forall p \in \mathcal{P}$.

$\bullet$ Finally, ratings vary in terms of {\it \textbf{reliability}}. 
This measure reflects how trustworthy the specific rating is.
The reliability $R(u,p)$ of a rating $(u,p)$ ranges from $0$ (an untrustworthy rating) to $1$ (a trustworthy rating) $\forall \texttt{(u,p)} \in \mathcal{R}$

\begin{figure}
    \centering
    \includegraphics[width=0.6\columnwidth, clip=True, trim=2mm 6mm 3mm 2mm]{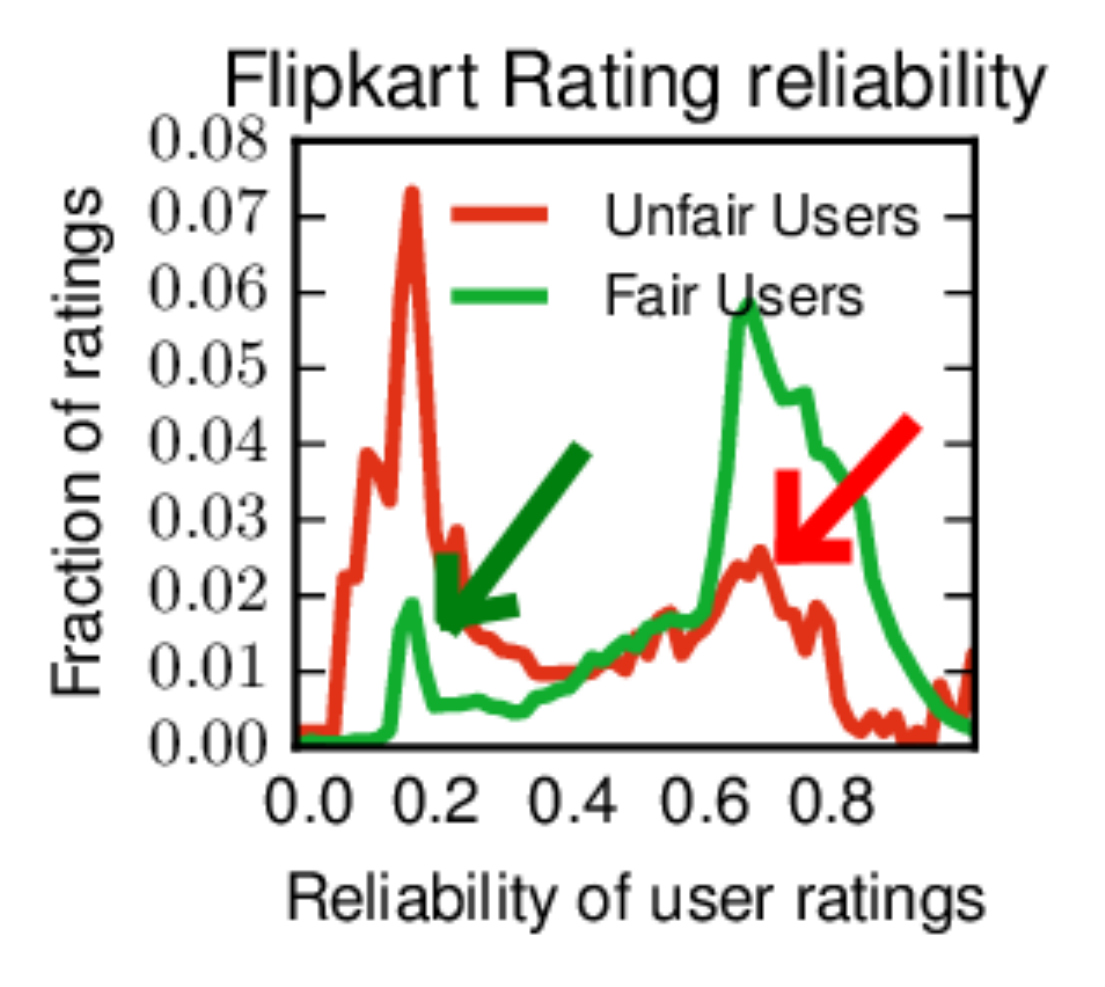} 
    \caption{While most ratings of fair users have high reliability, some ratings also have low reliability (green arrow). Conversely, unfair users also give some highly reliability ratings (red arrow), but most of their ratings have low reliability. \label{toy:reldist}}
            		\vspace{-4mm}

\end{figure}

The reader may wonder: {\em isn't the rating reliability, identical
to the user's fairness?} The answer is `no'. 
Consider Figure~\ref{toy:reldist}, where we show the rating reliability distribution of the top 1000 fair and top 1000 unfair users in the Flipkart network, as identified by our \methodall\ algorithm (explained later).
Notice that, while most ratings by fair users have high reliability, some of their ratings have low reliability, indicating personal opinions that disagree with majority (see green arrow). 
Conversely, unfair users give some high reliability ratings (red arrow), 
probably to camouflage themselves as fair users.
Thus, having reliability as a rating-specific metric allows us to more accurately characterize this distribution.

Given a bipartite user-product graph, we do not know the values of these three metrics for any user, product or rating. 
Clearly, these scores are mutually interdependent. 
Let us look at an example.

\begin{figure}[t]
    \begin{center}
           \includegraphics[width=0.8\columnwidth]{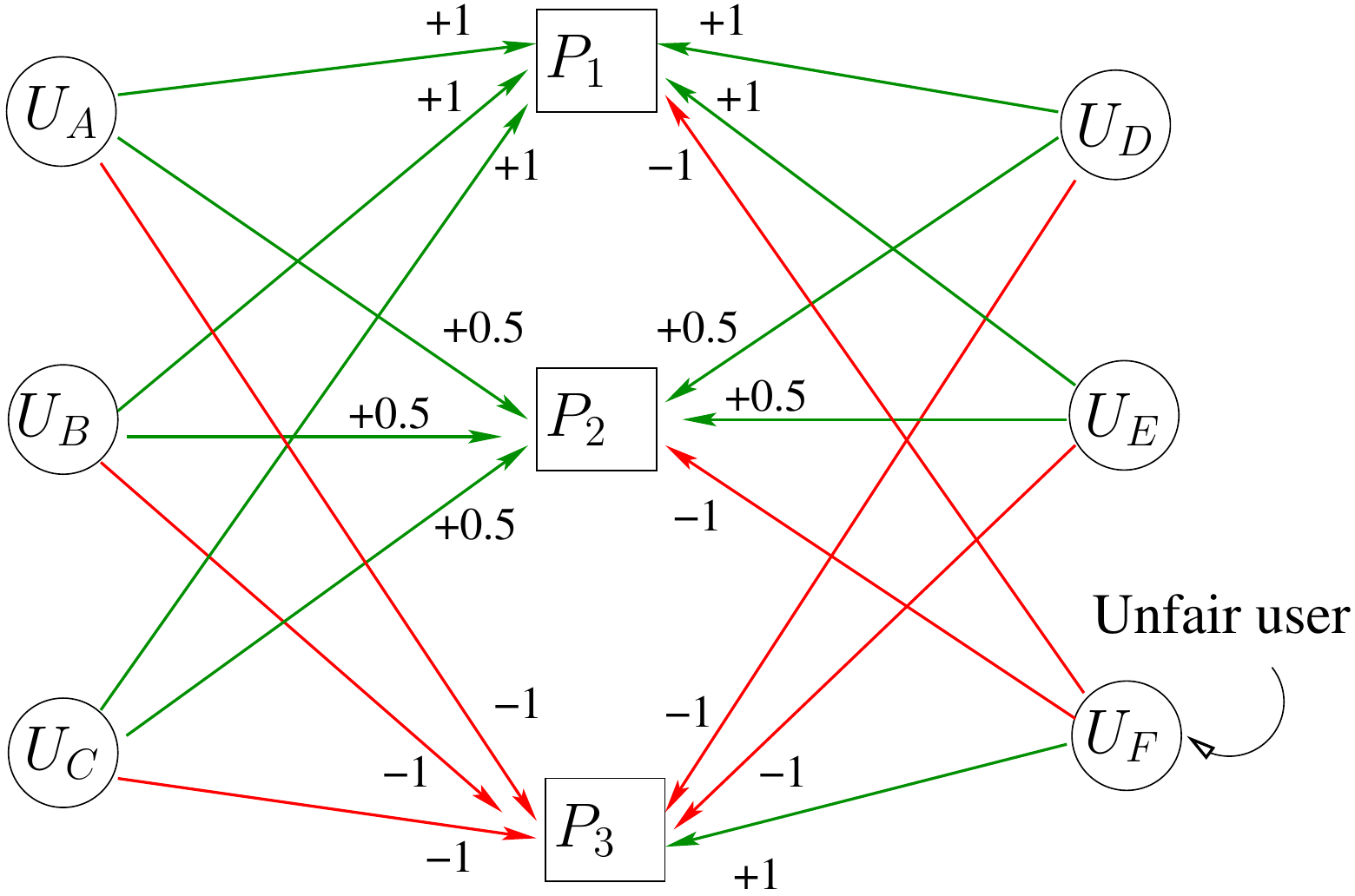} 
        \caption{Toy example showing products ($P_1$, $P_2$, $P_3$), users ($U_A$, $U_B$, $U_C$, $U_D$, $U_E$ and $U_F$), and rating scores provided by the users to the products. User $U_F$ always disagrees with the consensus, so $U_F$ is unfair.}
        		\vspace{-6mm}
        \label{toy:fig}
    \end{center}
\end{figure}

\begin{example}[Running Example]
Figure~\ref{toy:fig} shows a simple example in which there are 3 products, $P_1$ to $P_3$, and 6 users, $U_A$ to $U_F$.
Each review is denoted as an edge from a user to a product, with rating score between $-1$ and $+1$.
Note that this is a rescaled version of the traditional 5-star rating scale where a 1, 2, 3, 4 and 5 star corresponds to $-1, -0.5, 0, +0.5$ and $+1$ respectively. 

One can immediately see that $U_F$'s ratings are inconsistent with those of the $U_A, U_B$, $U_C$, $U_D$ and $U_E$. $U_F$ gives poor ratings to $P_1$ and $P_2$ which the others all agree are very good by consensus. $U_F$ also gives a high rating for $P_3$ which the others all agree is very bad. We will use this example to motivate our formal definitions below.
\end{example}

\begin{figure*}

\fbox{
\parbox{\textwidth}{
\begin{center}
\begin{eqnarray}
 \text{Fairness of user $u$, } F(u) & = & \frac{\hl{0.5 \cdot \alpha_1} +  \highlight{\alpha_2 \cdot IBIRDNEST_{IRTD_U}(u)} +  \sum_{(u,p) \in \text{\texttt{Out(u)}}} R(u,p)}{\hl{\alpha_1} + \highlight{\alpha_2} + |\text{\texttt{Out(u)}}|} \label{eqn:fairness_final}\\
 \text{ Reliability of rating $(u,p)$, } R(u,p) & = & \frac{1}{2}(F(u) + (1 - \frac{|\texttt{score(u,p)} - G(p)|}{2})) \label{eqn:reliability_final}\\
 \text{ Goodness of product $p$, } G(p) & = & \frac{ \highlight{\beta_2 \cdot IBIRDNEST_{IRTD_P}(p)} + \sum_{(u,p) \in \text{\texttt{In(p)}}} R(u,p) \cdot \texttt{score(u,p)}}{\hl{\beta_1} + \highlight{\beta_2} + |\text{\texttt{In(p)}}|} \hspace{8mm} \label{eqn:goodness_final}
 \end{eqnarray}
\end{center}
 }
 }
 \caption{This is the set of mutually recursive definitions of fairness, reliability and goodness for the proposed \methodall\ algorithm. The yellow shaded part addresses the cold start problems and gray shaded part incorporates the behavioral properties.}
 \label{eqn:basicFRG}
         		\vspace{-4mm}
 \end{figure*}

\xhdr{Fairness of users}
Intuitively, fair users are those who give reliable ratings, and unfair users mostly give unreliable ratings.
So we simply define a user's fairness score to be the average reliability score of its ratings:

\begin{align}
\label{eq:fairness}
F(u) = \frac{\sum\limits_{\texttt{(u,p)} \in \texttt{Out(u)}} R(u,p)}{|\texttt{Out(u)}|} 
\end{align}
Other definitions such as temporally weighted average can be developed, but we use the above for simplicity.

\xhdr{Goodness of products}
When a product receives rating scores via ratings with different reliability, clearly, more importance should be given to ratings that have higher reliability. 
Hence, to estimate a product's goodness, we weight the ratings by their reliability scores, giving higher weight to reliable ratings and little importance to low reliability ratings:

\begin{align}
\label{eq:goodness}
G(p) = \frac{\sum\limits_{\texttt{(u,p)} \in \texttt{In(p)}} R(u,p) \cdot \texttt{score(u,p)}}{|\texttt{In(p)}|}
\end{align}

Returning to our \textit{running example}, the ratings given by users $U_A$, $U_B$, $U_C$, $U_D$, $U_E$ and $U_F$ to product $P_1$ are $+1,+1,+1,+1,+1$ and $-1$, respectively. So we have:

\begin{align*}
G(P_1) = \frac{1}{6} (&R(U_A,P_1)+ R(U_B,P_1)+ R(U_C,P_1) + \\
& R(U_D,P_1) + R(U_E,P_1) - R(U_F,P_1))
\end{align*}

\xhdr{Reliability of ratings}
A rating $(u,p)$ should be considered reliable if (i) it is given by a generally fair user $u$, and (ii) its score is close to the goodness score of product $p$. The first condition makes sure that ratings by fair users are trusted more, in lieu of the user's established reputation. The second condition leverages `wisdom of crowds', by making sure that ratings that deviate from $p$'s goodness have low reliability. 
This deviation is measured as the normalized absolute difference, $|\texttt{score(u,p)} - G(p)| / 2$. Together:
\begin{align}
\label{eq:reliability}
R(u,p) = \frac{1}{2}(F(u) + (1 - \frac{|\texttt{score(u,p)} - G(p)|}{2}))
\end{align}

In our \textit{running example}, for the rating by user $U_F$ to $P_1$:
\begin{align*}
R(U_F,P_1) = \frac{1}{2}\left(F(U_F) + (1 - \frac{|-1 - G(P_1)|}{2})\right)
\end{align*}

Similar equations can be associated with every edge (rating) in the graph of Figure~\ref{toy:fig}.

\subsection{Addressing Cold Start Problems}
If a user $u$ has given 
only a few ratings, we have very little information about his true behavior. 
Say all of $u$'s ratings are very accurate -- it is hard to tell whether it is a fraudster that is camouflaging and building reputation by giving genuine ratings~\cite{hooi2016fraudar}, or it is actually a benign user. 
Conversely, if all of $u$'s ratings are very deviant, it is hard to tell whether the user is a fraudulent shill account~\cite{mukherjee2013yelp, lim2010detecting}, or simply a normal user whose rating behavior is unusual at first but stabilizes in the long run~\cite{birdnest}. 
Due to the lack of sufficient information about the ratings given by the user, little can be said about his fairness.
Similarly, for products that have only been rated a few times, it is hard to accurately determine their true quality, as they may be targets of fraud~\cite{lim2010detecting,trueview}.
This uncertainty due to insufficient information of less active users and products is the \textit{cold start problem}. 

We solve this issue by 
assigning Bayesian priors to each user's fairness score as follows:
\begin{equation}
F(u) = \frac{0.5 \cdot \alpha +  \sum_{(u,p) \in \text{\texttt{Out(u)}}} R(u,p)}{\alpha + |\text{\texttt{Out(u)}}|} \label{eqn:fairness_alpha}
\end{equation}

Here, $\alpha$ is a non-negative integer constant, which is the relative importance of the prior compared to the rating reliability -- the lower (higher) the value of $\alpha$, the more (less, resp.) the fairness score depends on the reliability of the ratings. 
The 0.5 score is the default global prior belief of all users' fairness, which is the midpoint of the fairness range $[0,1]$. 
If a user gives only a few ratings, then the fairness score of the user is close to the default score $0.5$. The more number of ratings the user gives, the more the fairness score moves towards the user's rating reliabilities. 
This way shills with few ratings have little effect on product scores.

Similarly, the Bayesian prior in product's goodness score is incorporated as:

\begin{equation}
G(p) = \frac{\sum_{(u,p) \in \text{\texttt{In(p)}}} R(u,p) \cdot \texttt{score(u,p)}}{\beta + |\text{\texttt{In(p)}}|} 
\label{eqn:goodness_alpha}
\end{equation}
Again, $\beta$ is a non-negative integer constant. The prior belief of product's goodness is set to $0$ which is the midpoint of the goodness range $[-1, 1]$, and so $0\cdot\beta = 0$ does not appear in the numerator. 
We will explain how the values of $\alpha$ and $\beta$ are set in Section~\ref{sec:algorithm}.

\subsection{Incorporating Behavioral Properties}
Rating scores alone are not sufficient to efficiently estimate the fairness, goodness and reliability values. 
The behavior of the users 
and products 
is also an important aspect to be considered.
As an example, fraudsters have been known to give several ratings in a very short timespan, or regularly at fixed intervals which indicates bot-like behavior~\cite{li2017bimodal,birdnest,viswanath2015strength}. 
Even though the ratings themselves may be very accurate, the unusually rapid behavior of the user is suspicious. 
Benign users, on the other hand, have a more spread out rating behavior as they lack regularity~\cite{li2017bimodal}.
In addition, products that receive an unusually high number of ratings for a very short time period may have bought fake ratings~\cite{wu2010merging,trueview}.
Therefore, behavioral properties of the ratings that a user gives or a product receives are indicative of their true nature.

Here, we show a Bayesian technique to incorporate the rating behavioral properties of users and products. 
We focus on temporal rating behavior as the behavioral property for the rest of the paper, but our method can be used with any additional set of properties. 
A user $u$'s temporal rating behavior is represented as the time difference between its consecutive ratings, \emph{inter-rating time distribution} $IRTD_U(u)$.
To model the behavior, we use a recent algorithm called BIRDNEST~\cite{birdnest}, which calculates a Bayesian estimate of how much user $u$'s $IRTD_U(u)$ deviates from the global population of all users' behavior, $IRTD_U(u') \forall u' \in \mathcal{U}$. 
This estimate is the BIRDNEST score of user $u$, $BIRDNEST_{IRTD_U}(u) \in [0,1]$. We use $IBIRDNEST_{IRTD_U}(u) = 1 - BIRDNEST_{IRTD_U}(u)$, as user $u$'s behavioral normality score. 
When \\ $IBIRDNEST_{IRTD_U}(u)$ = 1, it means the user's behavior is absolutely normal, and 0 means totally abnormal.
Similarly, for products, $IBIRDNEST_{IRTD_P}(p)$ is calculated from the consecutive ratings' time differences $IRTD_P(p)$ that product $p$ gets, with respect to the global $IRTD_P(p') \forall p' \in \mathcal{P}$.

As in the case of cold start, we adopt a Bayesian approach to incorporate the behavioral properties in the fairness and goodness equations~\ref{eqn:fairness_alpha} and ~\ref{eqn:goodness_alpha}. 
The $IBIRDNEST_{IRTD_U}(u)$ score is treated as user $u$'s behavioral Bayesian prior. 
As opposed to global priors (0.5 and 0) in the cold start case, this prior is user-dependent and may be different for different users. 
The resulting equation is given in Equation~\ref{eqn:fairness_final}.
Note that now there are two non-negative integer constants, $\alpha_1$ and $\alpha_2$, indicating the relative importance of the cold start and behavioral components, respectively, to the network component.
The Bayesian prior for products is incorporated similarly to give the final goodness equation in Equation~\ref{eqn:goodness_final}. 

Overall, the \methodall\ formulation represented in Figure~\ref{eqn:basicFRG} is the set of three equations that define the fairness, goodness and reliability scores in terms of each other. It combines the network, cold start treatment and behavioral properties together.

\section{The FairJudge Algorithm}
\label{sec:algorithm}
Having formulated the fairness, goodness and reliability metrics in Section~\ref{sec:formulation}, we present the \methodall\ algorithm in Algorithm~\ref{algo1} to calculate their values for all users, products and ratings.
The algorithm is iterative, so let $F^t, G^t$ and $R^t$ denote the fairness, goodness and reliability score at the end of iteration $t$. 
Given the rating network and non-negative integers $\alpha_1, \alpha_2, \beta_1$ and $\beta_2$,
we first initialize all scores to the highest value 1 (see line 3).\footnote{Random initialization gives similar results.}
Then we iteratively update the scores using Equations \eqref{eqn:goodness_final}, \eqref{eqn:reliability_final} and \eqref{eqn:fairness_final} until convergence (see lines 6-12). Convergence occurs when all scores change minimally (see line 12). $\epsilon$ is the acceptable error bound, which is set to a very small value, say $10^{-6}$.

But how do we set the values of $\alpha_1, \alpha_2, \beta_1$ and $\beta_2$? 
In an \textit{\textbf{unsupervised}} scenario, it is not possible to find the best combination of these values apriori. 
Therefore, the algorithm is run for several combinations of $\alpha_1, \alpha_2, \beta_1$ and $\beta_2$ as inputs, and 
the final scores of a user across all these runs are averaged to get the final \methodall\ score of the user.
Formally, let $\mathcal{C}$ be the set of all parameter combinations $\{\alpha_1, \alpha_2, \beta_1, \beta_2\}$, and $F(u | \alpha_1, \alpha_2, \beta_1, \beta_2)$ be the fairness score of user $u$ after running Algorithm~\ref{algo1} with $\alpha_1, \alpha_2, \beta_1$ and $\beta_2$ as input.
So the final fairness score of user $u$ is $F(u) = \frac{\sum_{(\alpha_1, \alpha_2, \beta_1, \beta_2) \in \mathcal{C}} F(u | \alpha_1, \alpha_2, \beta_1, \beta_2)}{|\mathcal{C}|}$. Similarly, \\
$G(p) = \frac{\sum_{(\alpha_1, \alpha_2, \beta_1, \beta_2) \in \mathcal{C}} G(p | \alpha_1, \alpha_2, \beta_1, \beta_2)}{|\mathcal{C}|}$ and \\
$R(u,p)$ $= \frac{\sum_{(\alpha_1, \alpha_2, \beta_1, \beta_2) \in \mathcal{C}} R((u,p) | \alpha_1, \alpha_2, \beta_1, \beta_2)}{|\mathcal{C}|}$.
In our experiments, we varied all (integer) parameters from 0 to 5, i.e. $0 \leq \alpha_1, \alpha_2, \beta_1, \beta_2 \leq 5$, giving 6x6x6x6 = 1296 combinations. 
So, altogether, scores from 1296 different runs were averaged to get the final \textit{unsupervised} \methodall\ scores.
This final score is used for ranking the users.

In a \textbf{\textit{supervised}} scenario, it is indeed possible to learn the relative importance of parameters. We represent each user $u$ as a feature vector of its fairness scores across several runs, i.e. $F(u | \alpha_1, \alpha_2, \beta_1, \beta_2), \forall (\alpha_1, \alpha_2, \beta_1, \beta_2) \in \mathcal{C}$ are the features for user $u$. Given a set of fraudulent and benign user labels, a random forest classifier is trained that learns the appropriate weights to be given to each score. The higher the weight, the more important the particular combination of parameter values is. The learned classifier's prediction labels are then used as the \textit{supervised} \methodall\ output.

{
\begin{algorithm}[t]
\small
\caption{\methodall\ Algorithm}
\label{algo1}
\begin{algorithmic}[1]
\STATE \textbf{Input}: Rating network $(\mathcal{U}, \mathcal{R}, \mathcal{P})$, $\alpha_1, \alpha_2, \beta_1, \beta_2$\\
\STATE \textbf{Output}: Fairness, Reliability and Goodness scores, given $\alpha_1, \alpha_2, \beta_1$ and $\beta_2$\\
\STATE Initialize $F^0(u)=1, R^0(u,p) = 1$ and $G^0(p)=1, \forall u \in \mathcal{U}, (u,p) \in \mathcal{R}, p \in \mathcal{P}$. \\
\STATE Calculate $IBIRDNEST_{IRTD_U}(u)$ $\forall u \in \mathcal{U}$ and $IBIRDNEST_{IRTD_P}(p)$  $\forall p \in \mathcal{P}$.\\
\STATE $t=0$
\STATE \textbf{do}\\
\STATE \ \  $t=t+1$\\
\STATE \ \  Update goodness of products using Equation~\ref{eqn:goodness_final}: $\forall p \in \mathcal{P},$ \\ 
\hspace{2mm}$G^t(p) = \frac{\beta_2 \cdot IBIRDNEST_{IRTD_P}(p) +  \sum_{(u,p) \in \texttt{In(p)}} R^{t-1}(u,p) . \texttt{score(u,p)}}{\beta_1 + \beta_2 + |\texttt{In(p)}|}$. \\ 
\STATE \ \  Update reliability of ratings using Equation~\ref{eqn:reliability_final} $\forall (u,p) \in \mathcal{R}$, \\ \hspace{2mm}$R^t(u,p) = \frac{1}{2}(F^{t-1}(u) + (1 - \frac{|\texttt{score(u,p)} - G^t(p)|}{2}))$. \\ 
\STATE \ \  Update fairness of users using Equation~\ref{eqn:fairness_final} $\forall u \in \mathcal{U}$, \\
\hspace{2mm}$ F^t(u) = \frac{0.5 \cdot \alpha_1 + \alpha_2 \cdot IBIRDNEST_{IRTD_U}(u) + \sum_{(u,p) \in \texttt{Out(u)}} R^t(u,p)}{\alpha_1 + \alpha_2 + |\texttt{Out(u)}|} $ \\ 
\STATE \ \  error = $\max(\max_{u \in \mathcal{U}} |F^{t}(u) - F^{t-1}(u)|$, $\max_{(u,p) \in \mathcal{R}}$ \\
\hspace{2mm} $|R^{t}(u,p) - R^{t-1}(u,p)|$, $\max_{p \in \mathcal{P}} |G^{t}(p) - G^{t-1}(p)|)$ \\
\STATE \textbf{while} $error > \epsilon $
\STATE \textbf{Return} $F^{t+1}(u)$, $R^{t+1}(u,p)$, $G^{t+1}(p)$, $\forall u \in \mathcal{U}, (u,p) \in \mathcal{R}, p \in \mathcal{P}$
\end{algorithmic}
\end{algorithm}
}

\begin{example}[Running Example]
Let us revisit our running example. 
Let, $\alpha_1 = \alpha_2 = \beta_1 = \beta_2 = 0$. We initialize all fairness, goodness and reliability scores to 1 (line 3 in Algorithm~\ref{algo1}. 
Consider the first iteration of the loop (i.e. when $t$ is set to 1 in line 7). In line 8, we must update the goodness of all products. Let us start with product $P_1$. Its goodness $G^0(P_1)$ was 1, but this gets updated to
{\small
\begin{eqnarray*}
G^1(P_1) & = & \frac{-1(1)+1(1)+1(1)+1(1)+1(1)+1(1)}{6}=0.67.
\end{eqnarray*}
}
We see that the goodness of $P_1$ has dropped because of $U_A$'s poor rating. The following table shows how the fairness and goodness values change over iterations (we omit reliability for brevity):
{
\small
\begin{center}
\begin{tabular}{|c|c|c|c|c|c|c|}
\hline
& & \multicolumn{5}{c|}{Fairness/Goodness in iterations} \\
Node & Property & 0 & 1 & 2 & 5 & 9 (final) \\\hline
$P_1$ & $G(P_1)$ & 1 & 0.67 & 0.67 & 0.67 & 0.68 \\
$P_2$ & $G(P_2)$ & 1 & 0.25 & 0.28 & 0.31 & 0.32 \\
$P_3$ & $G(P_3)$ & 1 & -0.67 & -0.67 & -0.67 & -0.68 \\\hline
$U_A$ - $U_E$ & $F(U_A)$ - $F(U_E)$ & 1 & 0.92 & 0.89 & 0.86 & 0.86 \\
$U_F$ & $F(U_F)$ & 1 & 0.62 & 0.43 & 0.24 & 0.22 \\\hline
\end{tabular}
\end{center}
}

By symmetry, nodes $U_A$ to $U_E$ have the same fairness values throughout. After convergence, $U_F$ has low fairness score, while $U_A$ to $U_E$ have close to perfect scores.
Confirming our intuition, the algorithm quickly learns that $U_F$ is an unfair user as all of its ratings disagree with the rest of the users. Hence, the algorithm then downweighs $U_F$'s ratings in its estimation of the products' goodness, raising the score of $P_1$ and $P_2$ as they deserve. 
\end{example}

\subsection{Theoretical Guarantees of FairJudge} \label{sec:theory}
Here we present the theoretical properties of \methodall.
Let, $F^\infty(u), G^\infty(p)$ and $R^\infty(u,p)$ be the final scores after convergence, for some input $\alpha_1, \alpha_2, \beta_1, \beta_2$.

\begin{lemma}[\textbf{Lemma 1}]
The difference between a product $p$'s final goodness score and its score after the first iteration is at most 1, i.e. $|G^{\infty}(p) - G^1(p)| \leq 1$. Similarly, $|R^{\infty}(u,p) - R^1(u,p)| \leq 3/4$ and $|F^{\infty}(u) - F^t(u)| \leq 3/4$.
\end{lemma}

The proof is shown in the appendix~\cite{appendix}. 

\begin{theorem}[\textbf{Convergence Theorem}]\label{th:conv}
The difference during iterations is bounded as $|R^{\infty}(u,p) - R^t(u,p)| \leq (\frac{3}{4})^t,$ $\forall (u,p) \in \mathcal{R}$. As $t$ increases, the difference decreases and $R^t(u,p)$ converges to $R^{\infty}(u,p)$.
Similarly, $|F^{\infty}(u) - F^t(u)| \leq (\frac{3}{4})^t, \forall u \in \mathcal{U}$ and $|G^{\infty}(p) - G^t(p)| \leq (\frac{3}{4})^{(t-1)}, \forall p \in \mathcal{P}$.
\label{thm:conv}
\end{theorem}

We prove this theorem formally in the appendix~\cite{appendix}.
As the algorithm converges for all $\alpha_1, \alpha_2, \beta_1, \beta_2$, the entire algorithm is guaranteed to converge.

\begin{corollary}[\textbf{Iterations till Convergence}]
The number of iterations needed to reach convergence is at most $2+\lceil \frac{\log(\epsilon/2)}{\log(3/4)} \rceil$. In other words, treating $\epsilon$ as constant, the number of iterations needed to reach convergence is bounded by a constant.
\label{thm:iterations}
\end{corollary}

Again, the proof is shown in the appendix~\cite{appendix}.

\xhdr{Linear algorithmic time complexity}
In each iteration, the algorithm updates goodness, reliability and fairness scores of each product, rating and user, respectively.
Each of these updates takes constant time.
So, the complexity of each iteration is $\mathcal{O}(|E| + |V|) = \mathcal{O}(|E|)$. 
By Corollary \ref{thm:iterations}, the algorithm converges in a constant number of iterations. Hence the time complexity is $\mathcal{O}(k|E|)$, which is linear in the number of edges, and $k$ is a constant equal to the product of the number of iterations till convergence and the number of runs of the algorithm.

\section{Experimental Evaluation}
\label{sec:exp}
In this section, we present the results of the proposed \methodall\ algorithm to identify fair and unfair users. 
We conduct extensive experiments on five different rating networks and show five major results:

\noindent (i) We compare \methodall\ algorithm with five recent algorithms to predict benign and malicious users in an unsupervised setting. We show that \methodall\ performs the best or second best in nine out of ten cases.

\noindent (ii) We show that \methodall\ outperforms nine algorithms across all datasets, when training data is available.
 
\noindent (iii) We show that \methodall\ is robust to the percentage of training data available, and consistently performs the best. 

\noindent (iv) We show that both cold start treatment and behavior properties improve the performance of \methodall\ algorithm, and incorporating both of them together perform the best.

\noindent (v) We show the linear running time of \methodall. 

The \methodall\ algorithm is already being deployed at Flipkart.

\subsection{Datasets: Rating Networks}
We use the following five datasets. Table~\ref{tab:datasets} has their properties. All ratings are rescaled between -1 and +1.\\
$\bullet$ \textbf{Flipkart} is India's biggest online marketplace where users rate products. 
The ground truth labels are generated by review fraud investigators in Flipkart,
who look at various properties of the user, rating and the product being rated.\\
$\bullet$ \textbf{Epinions} network has two components -- user-to-post rating network and 
user-to-user trust network~\cite{massa2007trust}.
Algorithms are run on the rating network and ground truth is defined using the trust network --
a user is defined as trustworthy if its total trust rating is $\geq +10$, 
and unfair if $\leq -10$.\\
$\bullet$ \textbf{Amazon} is a user-to-product rating network~\cite{mcauley2013amateurs}. 
The ground truth is defined using helpfulness votes, which is indicative of malicious behavior~\cite{fayazi2015uncovering} -- users with at least 50 votes are trustworthy (helpful) if the proportion of helpful-to-total votes is $\geq 0.75$,
and untrustworthy if $\leq 0.25$. \\
$\bullet$ \textbf{Bitcoin OTC} is a user-to-user trust network of Bitcoin users. 
The network is made bipartite by splitting each user into a `rater' with all its outgoing edges and `product' with all incoming edges.
The ground truth is defined as: trustworthy users are the platform's founder and users he rated highly positively ($\geq 0.5$).
Untrustworthy users are the ones that these trusted users give at least three more high negative ratings ($\leq -0.5$) than high positive ratings ($\geq 0.5$).\\
$\bullet$ \textbf{Bitcoin Alpha} is another Bitcoin trust network and its ground truth is created similar to OTC,
starting from the founder of this platform.

\begin{table}
\begin{center}
\small
\centering
\caption{Five rating networks used for evaluation.\label{tab:datasets}}
\vspace{-2mm}
\begin{tabular}{|c|lll|}
\hline
Network & \# Users (\% unfair, fair) &  \# Products & \# Edges \\\hline
OTC & 4,814 (3.7\%, 2.8\%) &  5,858 & 35,592 \\
Alpha & 3,286 (3.1\%, 4.2\%) & 3,754 & 24,186 \\
Amazon & 256,059 (0.09\%, 0.92\%)& 74,258 & 560,804  \\
Flipkart & 1,100,000 (-, -) & 550,000 & 3,300,000  \\
Epinions & 120,486 (0.84\%, 7.7\%) & 132,585 & 4,835,208 \\\hline
\end{tabular}

\vspace{-4mm}
\end{center}
\end{table}

\subsection{Baselines}
\label{sec:baselines}
We compare \methodall\ with nine state-of-the-art unsupervised and supervised algorithms.
The \textbf{unsupervised} algorithms are:\\
$\bullet$ \emph{Bias and Deserve (BAD)}~\cite{bad} assigns a bias score $bias(u)$ to each user $u$, which measures user's tendency to give high or low ratings.
$1-|bias(u)|$ is the prediction made by BAD.\\
$\bullet$ \emph{Trustiness}~\cite{icdm2011} algorithm assigns a trustiness, honesty and reliability score
to each user, product and rating, respectively. We use the trustiness score as its prediction.\\
$\bullet$ \emph{FraudEagle}~\cite{fraudeagle} is a belief propagation based algorithm. Users are ranked according to their fraud score.\\
$\bullet$ \emph{SpEagle}~\cite{speagle} incorporates behavior features into FraudEagle, and the final spam scores of users are used for ranking.\\
$\bullet$ \emph{BIRDNEST}~\cite{birdnest} ranks users by creating a Bayesian model with users' timestamp and rating distributions.

We also compare with \textbf{supervised} algorithms, when training labels are available:\\
$\bullet$ \emph{SpEagle+}~\cite{speagle} is a supervised extension of SpEagle that leverages known training labels in the ranking.\\
$\bullet$ \emph{SpamBehavior}~\cite{lim2010detecting}: This technique uses user's average rating deviations as feature.\\
$\bullet$ \emph{Spamicity}~\cite{mukherjee2013spotting} is creates each user's features as its review burstiness and maximum reviews per day.\\
$\bullet$ \emph{ICWSM'13}~\cite{mukherjee2013yelp} uses user's fraction of positive reviews, maximum reviews in a day and average rating deviation as features for prediction.\\

\begin{table*}
\caption{\underline{Unsupervised Predictions}: The table shows the Average Precision values of all algorithms in unsupervised prediction of unfair and fair users across five datasets. The {\color{blue!75}best algorithm} in each column is colored {\color{blue!75}blue} and {\color{gray!100}second best is gray}. Overall, \methodall\ performs the best or second best in 9 of the 10 cases. $nc$ indicates `no convergence'.\label{tab:unsupervised_results}
}
\begin{tabular}{|c||c|c|c|c|c||c|c|c|c|c|}
\hline
& \multicolumn{5}{c||}{Unfair user prediction} & \multicolumn{5}{c|}{Fair user prediction} \\
& OTC & Alpha & Amazon & Epinions & Flipkart & OTC & Alpha & Amazon & Epinions & Flipkart \\\hline
{\color{gray}FraudEagle} & \cellcolor{blue!25}93.67 & \cellcolor{blue!25}86.08 & 47.21 & $nc$ & $nc$ & \cellcolor{gray!25}86.94 & 71.99 & 96.88 & $nc$ & $nc$ \\
{\color{bad}BAD} & 79.75 & 63.29 & \cellcolor{gray!25}55.92 & \cellcolor{gray!25}58.31 & \cellcolor{gray!25}79.96 & 77.41 & 68.31 & 97.19 & 97.09 & 38.07 \\
{\color{speagle}SpEagle} & 74.40 & 68.42 & 12.16 & $nc$ & $nc$ & 80.91 & 82.23 & 93.42 & $nc$ & $nc$ \\
{\color{birdnest}BIRDNEST} & 61.89 & 53.46 & 19.09 & 37.08 & \cellcolor{blue!25}85.71 & 46.11 & 77.18 & 93.32 & \cellcolor{gray!25}98.53 & \cellcolor{blue!25}62.47 \\
{\color{trustiness}Trustiness} & 74.11 & 49.40 & 40.05 & $nc$ & $nc$ & 84.09 & \cellcolor{gray!25}78.19 & \cellcolor{gray!25}97.33 & $nc$ & $nc$ \\\hline
{\color{fairjudge}\textbf{FairJudge}} & \cellcolor{gray!25}86.03 & \cellcolor{gray!25}76.43 & \cellcolor{blue!25}56.18 &  \cellcolor{blue!25}63.43 & 57.14 & \cellcolor{blue!25}90.80 & \cellcolor{blue!25}86.16  & \cellcolor{blue!25}97.35 & \cellcolor{blue!25}99.35 & \cellcolor{gray!25}39.27 \\\hline
\end{tabular}
\vspace{-2mm}
\end{table*}

\vspace{-2mm}
\subsection{Experiment 1: Unsupervised Prediction}
In this experiment, the task is to rank the users based on their suspiciousness.
We compare \textit{unsupervised} \methodall\ with the suite of five unsupervised algorithms in terms of their \emph{Average Precision scores}, which measures the relative ordering the algorithm give to the known fair and unfair users. This score corresponds to the area under the precision-recall curve. We calculate two average precision scores -- one for fair users and another for unfair users.
Table~\ref{tab:unsupervised_results} shows the resulting average precision score on the five datasets. 

We see that \methodall\ performs the best in identifying fair users in 4 out of 5 networks, and second best in the Flipkart network. 
In identifying unfair users, our algorithm performs the best in two networks and second best in the two Bitcoin networks. 
The BIRDNEST algorithm is observed to perform quite well in case of Flipkart network, but has much weaker performance on other datasets. 
Note that FraudEagle, SpEagle and Trustiness are not scalable and do not converge for the two largest networks, Epinions and Flipkart, as opposed to \methodall\ which is guaranteed to converge. We discuss scalability in Section~\ref{sec:scalability}.

Overall, in unsupervised prediction, \methodall\ performs the best or second best in 9 out of 10 cases.

\begin{figure*}[t]
\centering
        \subfigure{
        \includegraphics[width=0.18\textwidth, clip=true, trim= 2.3mm 2.2mm 2.2mm 2.2mm]{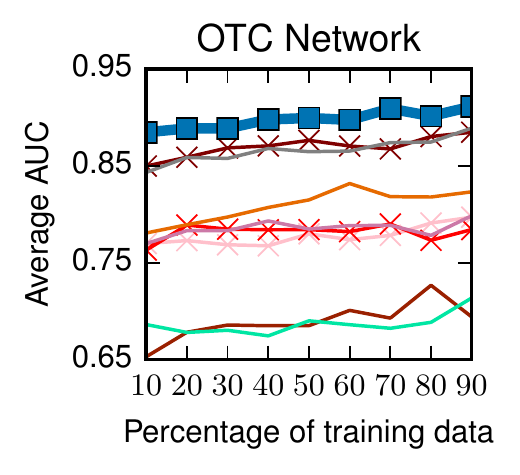}
        }
        \subfigure{
        \includegraphics[width=0.18\textwidth, clip=true, trim= 2.3mm 2.2mm 2.2mm 2.2mm]{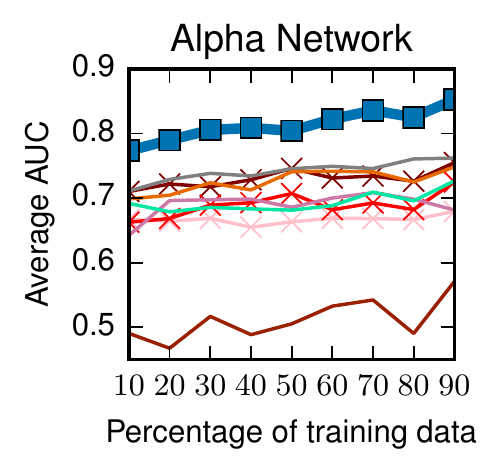}
        }
        \subfigure{
        \includegraphics[width=0.18\textwidth, clip=true, trim= 2.3mm 2.2mm 2.2mm 2.2mm]{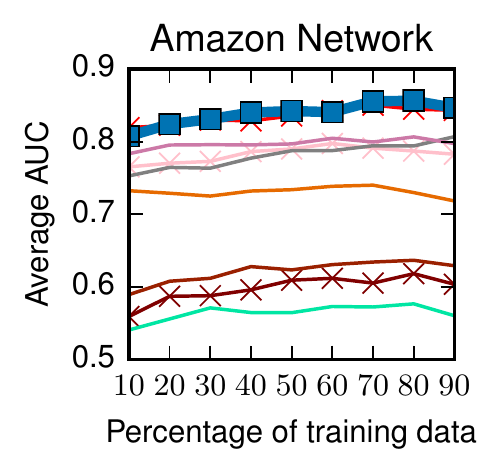}
        }
        \subfigure{
        \includegraphics[width=0.18\textwidth, clip=true, trim= 2.3mm 2.2mm 2.2mm 2.2mm]{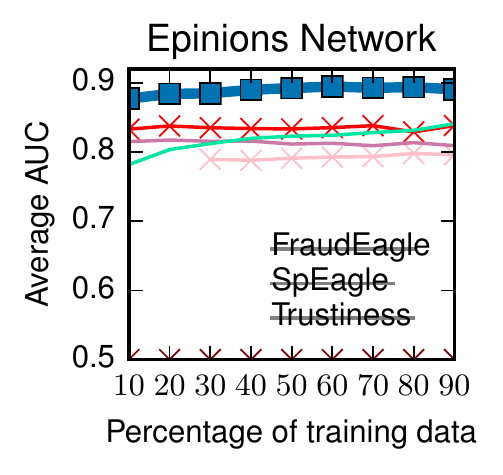}
        }
        \subfigure{
        \includegraphics[width=0.18\textwidth, clip=true, trim= 2.3mm 2.2mm 2.2mm 2.2mm]{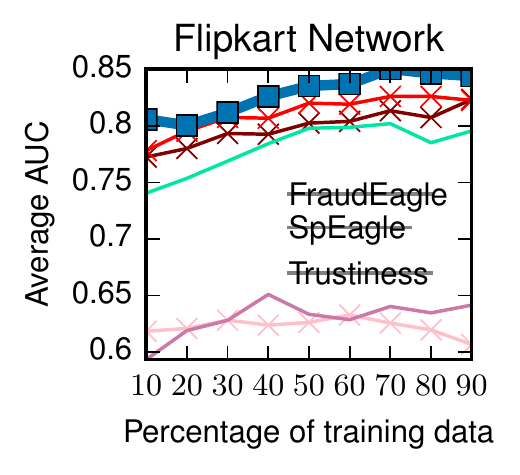}
        }
        
        \subfigure{
        \includegraphics[width=\textwidth, clip=true, trim= 2.3mm 1.7cm 2.2mm 5mm]{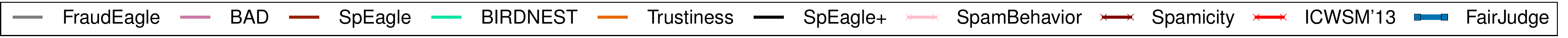}
        }
        \vspace{-4mm}
    \caption{\small Variation of AUC with percentage of training data available for supervision. \methodall\ consistently performs the best across all settings, and its performance is robust to the training percentage.}
    \label{fig:varying_N_results}
\vspace{-4mm}
\end{figure*}

\subsection{Experiment 2: Supervised Prediction}
In this experiment, the task is to predict the malicious and benign users, given some labels from both categories. 
The performance is measured using \emph{area under the ROC curve (AUC)} which is a standard measures when data is imbalanced, as is our case.
For each algorithm, a feature vector is created for each user and a binary classifier is trained.
As a reminder from Section~\ref{sec:algorithm}, for each user $u$, supervised \methodall\ creates a 1296 dimentional feature vector of its fairness scores $F(u | \alpha_1, \alpha_2, \beta_1, \beta_2)$, one for each of the 1296 combinations of $\alpha_1, \alpha_2, \beta_1, \beta_2 \in [0, 5]$.
For baselines FraudEagle, BAD, SpEagle, SpEagle+, BIRDNEST and Trustiness, 
the feature vector for user $u$ is the score the baseline gives to $u$ and $u$'s outdegree. 
For baselines SpamBehavior, Spamicity and ICWSM'13, 
the feature vector for user $u$ is their respective features given in Section~\ref{sec:baselines}.

We perform stratified 10-fold cross-validation using random forest classifier. The resulting AUCs are reported in Table~\ref{tab:supervised_results}. 
We see that \methodall\ outperforms all existing algorithms across all datasets and consistently has AUC $\geq 0.85$. 

Interestingly, supervised \methodall\ performs extremely well on the Flipkart dataset, while it did not perform as well on this dataset in the unsupervised experiment. 
This is because by using the training data, the classifier learns the importance of features $F(u | \alpha_1, \alpha_2, \beta_1, \beta_2)$ $\forall \{\alpha_1, \alpha_2, \beta_1, \beta_2\} \in \mathcal{C}$.
We reported the 100 most unfair users predicted by \methodall\ to review fraud investigators in Flipkart, and they found 80 users to be fraudulent (80\% accuracy).

\begin{table}
\vspace{0.5mm}
\caption{\underline{Supervised Predictions}: 10-fold cross validation with individual predictions as features in a Random Forest classifier. Values reported are AUC. \methodall\ performs the best across all datasets. $nc$ means `no convergence'. \label{tab:supervised_results}
}
\vspace{1mm}
\begin{tabular}{|c|c|c|c|c|c|}
\hline
& OTC & Alpha & Amazon & Epinions & Flipkart \\\hline
{\color{gray}FraudEagle} & \cellcolor{gray!25}0.89 & \cellcolor{gray!25}0.76 & 0.81 & $nc$ & $nc$ \\
{\color{bad}BAD} & 0.79 & 0.68 & 0.80 &  0.81 &  0.64\\
{\color{speagle}SpEagle} & 0.69 & 0.57 & 0.63 & $nc$ & $nc$ \\
{\color{birdnest}BIRDNEST} & 0.71 & 0.73 & 0.56 & \cellcolor{gray!25}0.84 & 0.80\\
{\color{trustiness}Trustiness} & 0.82 & 0.75 & 0.72 & $nc$ & $nc$ \\
SpEagle+ & 0.55 & 0.66 & 0.67 & $nc$ & $nc$ \\
{\color{sb}SpamBehavior} & 0.77 & 0.69 & 0.80 & 0.80 & 0.60\\
{\color{maroon}Spamicity} & 0.88 & 0.74 & 0.60 & 0.50 & \cellcolor{gray!25}0.82\\
{\color{red}ICWSM'13} & 0.75 & 0.71 & \cellcolor{gray!25}0.84 & 0.82 & \cellcolor{gray!25}0.82\\
\hline
{\color{fairjudge}\textbf{FairJudge}} & \cellcolor{blue!25}0.91 & \cellcolor{blue!25}0.85 & \cellcolor{blue!25}0.86 & \cellcolor{blue!25}0.89 & \cellcolor{blue!25}0.85 \\
\hline
\end{tabular}
\vspace{-5mm}
\end{table}

\subsection{Experiment 3: Robustness of FairJudge}
In this experiment, we evaluate the performance of the algorithms as the percentage of training data changes. 
We vary the training data from 10\% to 90\% in steps of 10.
Figure~\ref{fig:varying_N_results} shows the average AUC on test sets by using 50 random samples of training data.
We make two observations.
First, \methodall\ is robust to the amount of training data. Its performance is relatively stable (AUC $\geq 0.80$ in almost all cases) as the amount of training data varies. 
Second, \methodall\ outperforms other algorithms consistently across all datasets for almost all training percentages.
Together, these two show the efficiency of supervised \methodall\ algorithm even when small amount of training data is available.

\begin{figure}[t]
\centering
	\subfigure[Unsupervised]{
	\includegraphics[width=0.4\columnwidth, clip=true, trim= 2.3mm 2.2mm 2.2mm 2.2mm]{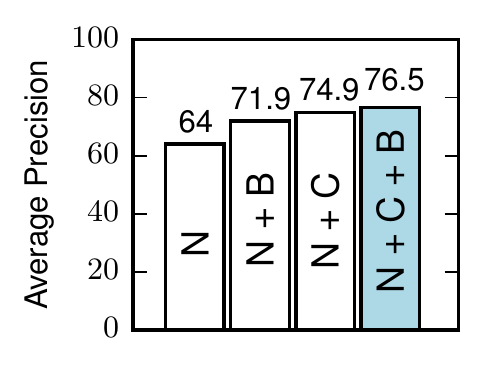}
	}
	\subfigure[Supervised]{
	\includegraphics[width=0.4\columnwidth, clip=true, trim= 2.3mm 2.2mm 2.2mm 2.2mm]{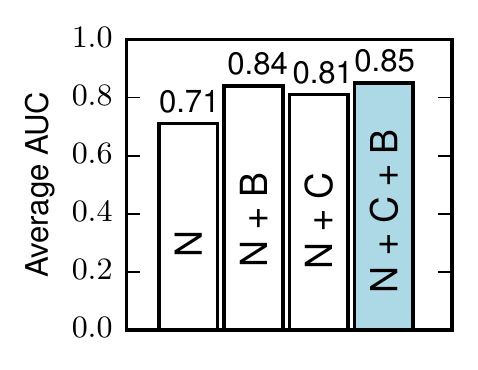}
	}
     \vspace{-2mm}
     \caption{\small Change in performance of \methodall\ on Alpha network in (a) unsupervised and (b) supervised experiments when different components are used: network (N), cold start treatment (C) and behavioral (B).}
	\vspace{-4mm}
    \label{fig:components}
\end{figure}

\subsection{Experiment 4: Importance of Network, Cold Start and Behavior}
In this experiment, we show the importance of the different components in the \methodall\ algorithm -- network (given by Equations~\ref{eq:fairness}, \ref{eq:goodness} and \ref{eq:reliability}; shown as N), cold start treatment (C), and behavioral properties (B). 
Figure~\ref{fig:components}(a) shows the average precision in unsupervised case for the Alpha dataset, when network property is combined with the other two components, and Figure~\ref{fig:components}(b) shows the average AUC in the supervised case. In both cases, network properties alone has the lowest performance. Adding either the cold start treatment component or the behavioral property component increases this performance. To combine network properties with behavioral property component only, the cold start property is removed by setting $\alpha_1 = \beta_1 = 0$ in the \methodall\ algorithm. Likewise, $\alpha_2$ and $\beta_2$ are set to 0 to combine with cold start treatment alone. 
Further, combining all three together gives the best performance. 
Similar observations hold for the other datasets as well.
This shows that all three components are important for predicting fraudulent users.

\begin{figure}[t]
\centering
	\subfigure[]{
	\includegraphics[width=0.4\columnwidth, clip=true, trim= 2.3mm 2.2mm 2.2mm 2.2mm]{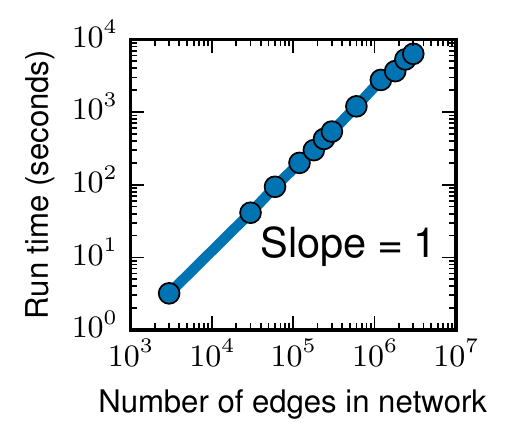}
	}
	\subfigure[]{
    \includegraphics[width=0.4\columnwidth, clip=true, trim= 2.3mm 2.2mm 2.2mm 2.2mm]{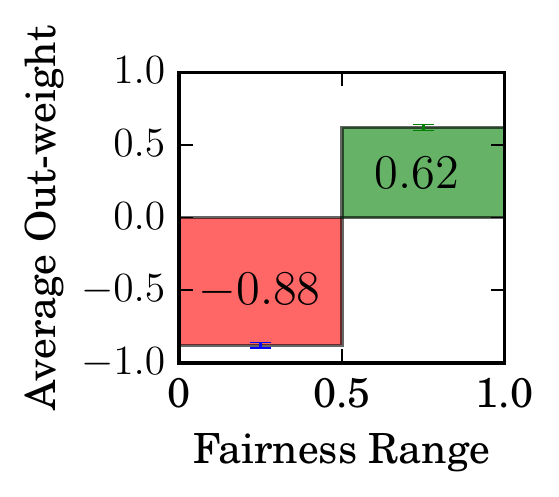}
    } 
     \vspace{-2mm}
     \caption{\small (a) \methodall\ scales linearly - the running time increases linearly with the number of edges. (b) Unfair users give highly negative ratings.}
\vspace{-6mm}
    \label{fig:scalability}
\end{figure}

\subsection{Experiment 5: Linear scalability}
\label{sec:scalability}
We have theoretically proved in Section~\ref{sec:theory} that \methodall\ is linear in running time in the number of edges. 
To show this experimentally as well, we create random networks of increasing number of nodes and edges and compute the running time of the algorithm till convergence.
Figure~\ref{fig:scalability} shows that the running time increases linearly with the number of edges in the network, which shows that \methodall\ algorithm is indeed scalable to large size networks for practical use.

\vspace{-1mm}
\subsection{Discoveries}

\begin{figure}[t]
\centering
		\subfigure[]{
        \includegraphics[width=0.45\columnwidth, clip=true, trim= 2.3mm 2.2mm 2.2mm 2.2mm]{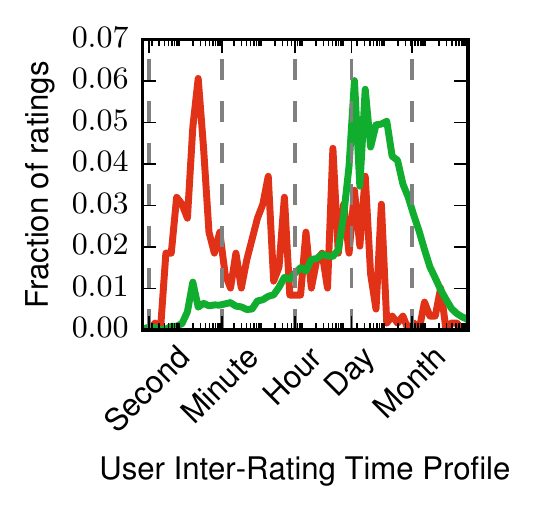}
        }
        \subfigure[]{
        \includegraphics[width=0.45\columnwidth, clip=true, trim= 2.3mm 2.2mm 2.2mm 2.2mm]{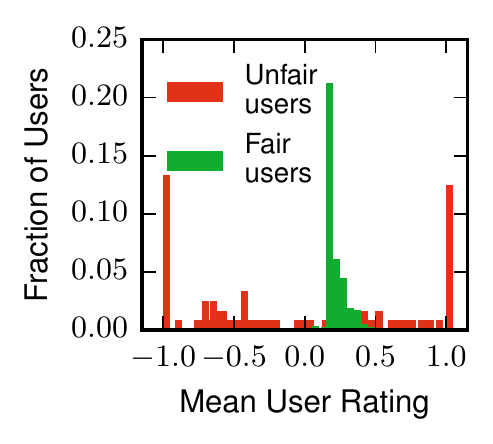}
        \label{fig:btg}
        }
        \vspace{-2mm}
    \caption{\small Identified unfair users by \methodall\ are (a) faster in rating , and give extreme ratings.}
		\vspace{-8mm}
    \label{fig:otc-properties}
\end{figure}

Here we look at some insights and discoveries about malicious behavior found by \methodall.

As seen previously in Figure~\ref{toy:reldist}, most ratings given by unfair users have low reliability, while some have high reliability, indicating camouflage to masquerade as fair users. 
At the same time, most ratings of fair users have high reliability, but some ratings have low reliability, indicating personal opinion about products that disagrees with `consensus'.

We see in Figure~\ref{fig:scalability}(b) that in the Amazon network, unfair users detected by \methodall\ ($F(u) \leq 0.5$) give highly negative rating on average (mean out-rating weight of -0.88), while fair users give positive ratings (mean out-weight 0.62).
Unfair users also give almost half as many ratings as fair users (7.85 vs 15.92).
This means on average, fraudsters aggressively bad-mouth target products. 

We also observe in Figure~\ref{fig:otc-properties} that unfair users in OTC network: (a) give ratings in quick succession of less than a few minutes, and (b) exhibit bimodal rating pattern -- either they give all -1.0 ratings (possibly, bad-mouthing a competitor) or all +1.0 ratings (possibly, over-selling their products/friends). As an example, the most unfair user found by \methodall\ had 3500 ratings, all with +0.5 score, and almost all given 15 seconds apart (apparently, a script).
On the other hand, fair users have a day to a month between consecutive ratings, and they give mildly positive ratings (between 0.1 and 0.5).
These observations are coherent with existing research~\cite{li2017bimodal,cheng2015antisocial}.

Figure~\ref{fig:crownjewel}(d) shows a set of 40 unfair users, as identified by \methodall\ on the Alpha network, that collude to positively rate each other and form sets of tightly knit clusters -- 
they are confirmed to be shills of a single user.

In summary, unfair users detected by \methodall\ exhibit 
strange characteristics with respect to their behavior:\\
$\bullet$ They have bimodal rating pattern - they give too low (bad-mouthing) or too high (over-selling) ratings. \\
$\bullet$ They are less active, have no daily periodicity, and post quickly, often less than a few minutes apart. \\
$\bullet$ They tend to form near-cliques, colluding to boost their own or their product's ratings.\\
$\bullet$ They camouflage their behavior to masquerade as benign users.

\section{Conclusions}
\label{sec:concl}
We presented the \methodall\ algorithm to address the problem of identifying fraudulent users in rating networks.
This paper has the following contributions:\\
$\bullet$ \textbf{Algorithm}: We proposed three mutually-recursive metrics - fairness of users, goodness of products and reliability of ratings. We extended the metrics to incorporate Bayesian solutions to cold start problem and behavioral properties. 
We proposed the \methodall\ algorithm to iteratively compute these metrics.\\
$\bullet$ \textbf{Theoretical guarantees}: We proved that \methodall\ algorithm has linear time complexity and is guaranteed to converge in a bounded number of iterations. \\ 
$\bullet$ \textbf{Effectiveness}: By conducting five experiments, we showed that \methodall\ outperforms nine existing algorithms to predict fair and unfair users. \methodall\ is practically useful, and already under deployment at Flipkart.

\bibliographystyle{abbrv}
\bibliography{bibliography}

\end{document}